\documentclass[sigconf]{acmart}

\usepackage{url}
\usepackage[dvipsnames, svgnames, x11names]{xcolor} % 一般放在最前面
\usepackage{xcolor}
\usepackage{enumitem}
\usepackage{xspace}
\usepackage{makecell}
\usepackage{multirow}
\usepackage{pifont}
\usepackage{listings}
\usepackage{setspace}
\usepackage[lined,boxed,commentsnumbered]{algorithm2e}
\usepackage{color}
\usepackage{graphicx}
\usepackage{listings}
\usepackage{tikz}
\usepackage{cancel}
\usetikzlibrary{positioning}
\usepackage{subcaption}

\def\Snospace~{\S{}}

\newcommand{\vulnsage}{\mbox{\texttt{VulnSage}}\xspace}
\newcommand{\nodemedicfine}{\mbox{\texttt{\path{NodeMedic-FINE}}}\xspace}
\newcommand{\explodejs}{\mbox{\texttt{\path{EXPLOADE.js}}}\xspace}
\newcommand{\validset}{\mbox{\textit{Real-world Dataset}}\xspace}
\newcommand{\secbench}{\mbox{\textit{SecBench.js}}\xspace}
\newcommand{\vulnrelated}{\textit{alert information}\xspace}

\newcommand{\commandinjection}{{CmdInj.}\xspace}
\newcommand{\codeinjection}{{CodeInj.}\xspace}
\newcommand{\jndiinjection}{{JNDI.}\xspace}
\newcommand{\pathtraversal}{{Path.}\xspace}
\newcommand{\prototypepollution}{{Proto.}\xspace}

\newcommand{\code}[1]{\texttt{\path{#1}}}
\newcommand\etal[1]{#1 \textit{et al.}}
\newcommand{\callchain}{\textit{\path{callChainWithCtx\xspace}}}

\definecolor{dgreen}{rgb}{0.25,0.5,0.40}
\definecolor{black}{rgb}{0.0,0.0,0.0}
\newcommand\num[1]{{\textcolor{black}{#1}}}

\definecolor{javagreen}{rgb}{0.25,0.5,0.35}
\definecolor{javapurple}{rgb}{0.5,0,0.35}
\definecolor{javared}{rgb}{0.6,0,0}
\definecolor{linenumcolor}{rgb}{0.1,0.1,0.1}
\lstdefinelanguage{JavaScript}{
  keywords={break, case, catch, continue, debugger, default, delete, do, else, false, finally, for, function, if, in, instanceof, new, null, undefined, NaN, return, switch, this, throw, true, try, typeof, var, void, while, with, let, module, export, import, from},
  morecomment=[l]{//},
  morecomment=[s]{/*}{*/},
  morestring=[b]',
  morestring=[b]",
  ndkeywords={class, export, boolean, throw, implements, import, this},
    commentstyle=\color{javagreen},
    keywordstyle=\color{blue},
    stringstyle=\color{javared},
  ndkeywordstyle=\color{darkgray}\bfseries,
  identifierstyle=\color{black},
  escapeinside={(*@}{@*)},
  sensitive=true,
  xleftmargin=2em,xrightmargin=2em, aboveskip=1em,
  framexleftmargin=2em
}

\copyrightyear{2026}
\acmYear{2026}
\setcopyright{cc}
\setcctype{by}
\acmConference[ICPC '26]{34th IEEE/ACM International Conference on Program Comprehension}{April 12--13, 2026}{Rio de Janeiro, Brazil}
\acmBooktitle{34th IEEE/ACM International Conference on Program Comprehension (ICPC '26), April 12--13, 2026, Rio de Janeiro, Brazil}
\acmPrice{}
\acmDOI{10.1145/3794763.3794817}
\acmISBN{979-8-4007-2482-4/2026/04}

\setcopyright{none}
\begin{document}

\title{A Multi-Agent Framework for Automated Exploit Generation with Constraint-Guided Comprehension and Reflection}

\settopmatter{authorsperrow=4}

\author{Siyi Chen}
\authornote{co-first authors}
\affiliation{%
  \institution{Alibaba Group}
  \city{Hangzhou}
  \country{China}
}
\email{siyi.csy@alibaba-inc.com}

\author{Tianhan Luo}
\authornotemark[1]
\affiliation{%
  \institution{Alibaba Group}
  \city{Hangzhou}
  \country{China}
}
\email{luotianhan.lth@alibaba-inc.com}

\author{Shijian Wu}
\affiliation{%
  \institution{Alibaba Group}
  \city{Hangzhou}
  \country{China}
}
\email{wushijian.wsj@alibaba-inc.com}

\author{Xiangyu Liu}
\affiliation{%
  \institution{Alibaba Group}
  \city{Hangzhou}
  \country{China}
}
\email{eason.lxy@alibaba-inc.com}

\author{Yilin Zhou}
\affiliation{%
  \institution{Wuhan University}
  \city{Wuhan}
  \country{China}
}
\email{ylzhou@whu.edu.cn}

\author{Qi Li}
\affiliation{%
  \institution{Alibaba Group}
  \city{Hangzhou}
  \country{China}
}
\email{insomia.liq@taobao.com}

\author{Wenyuan Xu}
\affiliation{%
  \institution{Aarhus University}
  \city{Aarhus}
  \country{Denmark}
}
\email{wenyuan.xu@cs.au.dk}




\begin{abstract}

Open-source libraries are widely used in modern software development, introducing significant security vulnerabilities. While static analysis tools can identify potential vulnerabilities at scale, they often generate overwhelming reports with high false positive rates. Automated Exploit Generation (AEG) emerges as a promising solution to confirm vulnerability authenticity by generating an exploit. 
However, traditional AEG approaches based on fuzzing or symbolic execution face path coverage and constraint-solving problems.
Although LLMs show great potential for AEG, how to effectively leverage them to comprehend vulnerabilities and generate corresponding exploits is still an open question.

To address these challenges, we propose \vulnsage, a multi-agent framework for AEG.
\vulnsage simulates human security researchers' workflows by decomposing the complex AEG process into multiple specialized sub-agents: Code Analyzer Agent, Code Generation Agent, Validation Agent, and a set of Reflection Agents, orchestrated by a central supervisor through iterative cycles.
Given a target program, the Code Analyzer Agent performs static analysis to identify potential vulnerabilities and collects relevant information for each one. 
The Code Generation Agent then utilizes an LLM to generate candidate exploits. 
The Validation Agent and Reflection Agents form a feedback-driven self-refinement loop that uses execution traces and runtime error analysis to either improve the exploit iteratively or reason about the false positive alert.

Experimental evaluation demonstrates that \vulnsage succeeds in generating 34.64\% more exploits than state-of-the-art tools such as \explodejs. Furthermore, \vulnsage has successfully discovered and verified 146 zero-day vulnerabilities in real-world scenarios, demonstrating its practical effectiveness for assisting security assessment in software supply chains.

\end{abstract}

\maketitle

\section{Introduction}

The widespread use of open-source libraries in modern software ecosystems introduces significant security risks~\cite{DBLP:journals/tse/ManesHHCESW21,pashchenko2024vulnerability}. As a result, automated discovery and mitigation of library vulnerabilities have become an important research focus in cybersecurity~\cite{koo2024uncovering,google2024ossfuzz}.

Static analysis has been widely adopted for automated vulnerability detection. 
However, traditional static analysis methods often struggle to achieve scalability, soundness, and precision simultaneously~\cite{goseva2015capability}. Many practical tools, such as CodeQL~\cite{codeql_docs}, prioritize scalability and soundness, but this trade-off often leads to overwhelming reports with numerous false positives, requiring significant manual effort for verification.

Automated Exploit Generation (AEG) has emerged as a promising solution to this problem, aiming to automatically generate a working proof-of-concept (PoC) to confirm a vulnerability's authenticity and impact.
Despite its advantages, AEG remains limited in real-world exploit validation.
This is primarily due to the difficulty of crafting inputs required by the entry point function~\footnote{In our scenario, the function that contains user-controlled parameters is the entry point function.}. 
Such inputs often have complex data structures, and their values must satisfy specific control-flow constraints in order to reach the critical function (i.e., sink in taint analysis terminology) and finally execute attacker-defined payload~\cite{avgerinos2011aeg,cha2012mayhem}.

Traditional dynamic analysis techniques, such as fuzzing, have demonstrated success in discovering vulnerabilities by providing unexpected inputs~\cite{zalewski2014american, bohme2016aflfast,bohme2017directed}.
However, their effectiveness diminishes when faced with complex program logic or deep execution paths. Another group of techniques using symbolic execution, employed by tools such as \explodejs~\cite{marques2025explode}, explores the program paths by translating them into formal constraints.
However, the constraints encoded in code, including string manipulations and higher-order functions, are complex and difficult for current SMT frameworks to solve effectively~\cite{baldoni2018survey}.

Recent advancements in Large Language Model (LLM) show their powerful code understanding and imitation capabilities, which could be great potential for code generation tasks~\cite{lutellier2023large, nunez2024multi}. 
However, applying LLM in AEG also has challenges~\cite{2505.01065}: (1) the LLM only produces useful outputs when it fully understands the task. For generating an exploit for a vulnerability, the LLM must firstly comprehend all code relevant to the vulnerability, which often exceeds a single model's context window~\cite{DBLP:conf/coling/HosseiniCGP25}~\cite{DBLP:journals/corr/abs-2505-21471}.
(2) Even when the code fits the context window, large language models may lose focus midway and produce hallucinated content~\cite{liu2023lost,zhang2024comprehensive,wang2024longcontext}. 
(3) AEG is a complex task, including code analysis, code synthesis, exploit verification, etc. --- without special handling, LLMs struggle with complex or multi-step tasks~\cite{DBLP:conf/nips/Wei0SBIXCLZ22}.
(4) LLM can not verify the generated code itself, nor the correctness of the code's syntax.

To address the limitations above, we propose \vulnsage, a multi-agent framework for automated exploit generation.
\vulnsage is guided by a key insight: An experienced security analyst would break down the exploit process into four distinct steps:
(1) auditing the vulnerable code deeply and comprehending the input structure, 
(2) writing exploit code, 
(3) executing the exploit to check whether it is successful, 
(4) when execution fails, adjusting the exploit with feedback.
These four complementary subtasks require multiple iterations to achieve successful exploitation. 
Accordingly, \vulnsage incorporates several specialized agents: Code Analyzer Agent, Code Generation Agent, Validation Agent, and a set of Reflection Agents. 
These agents are invoked by a Supervisor Agent based on the current progress of exploit generation to simulate the human AEG workflow.
Given a library, the Code Analyzer Agent performs static analysis, 
extracts vulnerability-related code slices for each alert, and drafts an initial exploit template. 
The Code Generation Agent executes a two-step process: (1) encoding the exploit generation problem as a set of constraints and (2) generating exploit code that satisfies those constraints.
The Validation Agent prepares an executable environment for exploits and verifies exploit validity using oracles.
The Reflection Agents include an agent that derives new constraints from failed exploits and an agent that reasons whether the corresponding alert is a false positive.
Finally, the Supervisor Agent outputs the confirmed vulnerabilities with their corresponding exploits, along with identified false positives and the underlying reasons.

Our experimental results demonstrate that \vulnsage successfully generates exploits for 53.47\% vulnerabilities on \secbench, a significant improvement of at least 34.64\% over the state-of-the-art tools such as \explodejs on JavaScript vulnerabilities. Furthermore, \vulnsage successfully discovered \num{146} 0-day vulnerabilities with reasonable cost.

\noindent
\textbf{Contributions.}The paper’s contributions can be briefly outlined as follows:
\begin{enumerate}[leftmargin=0.5cm, itemindent=0cm,topsep=0cm]
\item

\textbf{Multi-Agent Vulnerability Discovery Architecture:} We propose a novel multi-agent framework named \vulnsage, addressing the context size limitation of large language models in the AEG scenario.

\item \textbf{Constraints-based Comprehension:} We extract the relevant code and encode it as constraints to guide the LLM toward a better understanding of the vulnerability.

\item
\textbf{Environment Feedback and Reflection Mechanism:} We develop a continuous reflection mechanism that can automatically validate the exploits, leverage the feedback to refine the exploit code, and reason about false positives of alerts.

\item
\textbf{Empirical Effectiveness:} 
We conduct experimental evaluations that demonstrate superior performance compared to state-of-the-art baselines, with successful discovery of multiple zero-day vulnerabilities in real-world scenarios.

\end{enumerate}

\section{Motivating Example}\label{sec:MOTIVATING}

To illustrate the core challenges in automated exploit generation (AEG), we present a real-world Java Naming and Directory Interface (JNDI) injection vulnerability, \href{https://nvd.nist.gov/vuln/detail/CVE-2023-39017}{CVE-2023-39017}, in \autoref{lst:soure_code_cve}.

The entry point is \code{SendMsgJob.execute}(line~\ref{line:sendmsg.execute}), which accepts a user-controllable parameter \code{jobCtx}. 
\code{jobCtx} contains a \code{JobDataMap} that holds an entry named \code{jms.connection.factory}; 
this entry is retrieved via the calls on line~\ref{line:sendmsg.dataMap} and line~\ref{line:sendmsg.connection}. 
Finally, the method performs a JNDI lookup (line~\ref{line:sendmsg.lookup}).
An attacker who controls \code{jms.connection.factory} can therefore supply a malicious JNDI URI (e.g., \url{ldap://evil.example.com/Foo}) and cause remote code loading and execution.

\begin{figure}[H]
    \centering
        \begin{lstlisting}
package org.quartz.jobs.ee.jms;
class SendMsgJob extends Job {
    public void execute(JobExecCtx jobCtx){ (*@\label{line:sendmsg.execute}@*)
    JobDataMap dataMap = jobCtx.getJobDataMap(); (*@\label{line:sendmsg.dataMap}@*)
    Context namingCtx =JmsHelper.getContext(dataMap);
    namingCtx.lookup( (*@\label{line:sendmsg.lookup}@*)
        dataMap.getString("jms.connection.factory")); (*@\label{line:sendmsg.connection}@*)
    ...}
}
class JobExecCtxImpl implements JobExecCtx{
    public JobExecCtx(JobDetail jobDetail, Trigger trigger) { (*@\label{line:JobExecTexImpl}@*)
    this.jobDataMap = jobDetail.getJobDataMap();  (*@\label{line:JobExecCtxImpl.jobDetail}@*)
    this.jobDataMap.putAll(trigger.getJobDataMap());(*@\label{line:JobExecCtxImpl.trigger}@*)
    ...}
}
class JobDetailImpl implements JobDetail{
    public JobDetail(Class<? extends Job> jobClass) { (*@\label{line:JobDetailImpl}@*)
    this.jobClass = jobClass;
    ...}
}
class BaseTriggerImpl implements Trigger{
    public BaseTrigger() {
        this.jobDataMap = new JobDataMap();
        this.jobDataMap.put("jms.connection.factory", "ldap://benign/uri");
        (*@\label{line:BaseTrigger.put}@*)
    }
}
class SimlpeTriggerImpl implements Trigger{
    public SimlpeTrigger() {}
}
        \end{lstlisting}
        \vspace{-0.15cm}
    \caption{Source code snippet of CVE-2023-39017.}
    \label{lst:soure_code_cve}
    \vspace{-1em}
\end{figure}

A successful exploit code generated by \vulnsage is shown in \autoref{lst:success_exp}.
The main challenge of generating such exploit code is preparing the input to call entry point functions.
To reach the \code{namingCtx.lookup} (line~\ref{line:sendmsg.lookup}, \autoref{lst:soure_code_cve}), the exploit code must ensure that the input controls the exploit execution as the attacker intended to --- executing malicious JNDI look up from URL in line~\ref{line:exploit.ldap}, \autoref{lst:success_exp}.
However, constructing \code{jobCtx} is hard.
According to the constructor of \code{JobExecTexImpl}(line~\ref{line:JobExecTexImpl}, \autoref{lst:soure_code_cve}), instantiating \code{jobCtx} requires constructing a \code{JobExecCtx} object with two non-null arguments: 
\code{jobDetail} and \code{trigger}. Null arguments will result in \code{NullPointerException} before reaching the vulnerability due to lines \ref{line:JobExecCtxImpl.jobDetail}-\ref{line:JobExecCtxImpl.trigger}, \autoref{lst:soure_code_cve}. 
According to the constructor of \code{JobDetailImpl} (line~\ref{line:JobDetailImpl}, \autoref{lst:soure_code_cve}), instantiating \code{JobDetail} requires an appropriate class-type.
Although the class-type is parameterized by a generic type \code{Class<? extends Job>}, in the real code base, the abstract class \code{Job} has 13 sub-classes as candidates.
Furthermore, although we only showed two subclasses implementing Trigger. 
There are 12 subclasses in real code.
Choosing the wrong candidates results in an unsuccessful exploit.
For example, in line~\ref{line:exploit.JobExecCtxImpl}, \autoref{lst:success_exp}, if we choose \code{BaseTriggerImpl} instead, the function will not take the attacker-controlled URI because the benign URI of \code{jobDataMap} defined in \code{BaseTrigger}(line~\ref{line:BaseTrigger.put}, \autoref{lst:soure_code_cve}) overrides the malicious URI in \code{JobExecCtx}(line \ref{line:JobExecCtxImpl.trigger}, \autoref{lst:soure_code_cve}).

\begin{figure}[t]
    \centering
\begin{lstlisting}
import org.quartz.impl.triggers.SimpleTriggerImpl;
import org.quartz.jobs.ee.jms.SendMsgJob;
import org.quartz.impl.*;
...
JobDataMap dataMap = new JobDataMap();
dataMap.put("jms.connection.factory", 
            "ldap://127.0.0.1:1099/Foo"); (*@\label{line:exploit.ldap}@*)
JobDetail job = new JobDetailImpl(SendMsgJob.class);
job.setJobDataMap(dataMap);
SimpleTriggerImpl trigger = new SimpleTriggerImpl();
JobExecCtxImpl ctx =new JobExecCtxImpl(trigger, job); (*@\label{line:exploit.JobExecCtxImpl}@*)
SendMsgJob instance = new SendMsgJob();
instance.execute(ctx);
\end{lstlisting}
\vspace{-0.15cm}
    \caption{Successful exploit of CVE-2023-39017 by \vulnsage.}
    \label{lst:success_exp}
    \vspace{-1em}
\end{figure}

Polymorphism patterns, as detailed above, are widely used in object-oriented languages like Java and JavaScript, leading to an exponential search space for AEG.
Fuzzing-based methods\cite{bohme2016aflfast, zalewski2014american} struggle to generate a valid \code{JobExecCtx} instance with appropriate constructor parameters.
Symbolic execution approaches~\cite{avgerinos2011aeg, stephens2016driller} also face fundamental limitations: no existing SMT can solve constraints of the form ``a function (or class) that satisfies certain properties'' as such reasoning requires higher-order logic.
Furthermore, the presence of higher-order functions and complex string operations --- as we will demonstrate in \autoref{sec:evaluation} --- further increases the difficulties of AEG.

LLM might solve the problem from a completely different perspective.
It has seen billions of lines of code during the training phase.
Rather than attempting to solve the hard constraints directly, it generates the realistic code for calling the entry point functions by mimicking how the real code would call the function. 
Since the code that LLMs learn from is written by real-world developers, it naturally satisfies the invocation conditions.
As a result, the code generated by LLMs is also likely to satisfy those constraints without explicitly solving them. 

However, LLMs have their own limitations. Even if we temporarily ignore the context-length restriction, other limitations such as hallucination and attention degradation cause GPT-4o to generate an unsuccessful exploit, as shown in \autoref{lst:fail_expl}.

\begin{figure}[t]
    \centering
\begin{lstlisting}
import org.quartz.impl.triggers.BaseTriggerImpl;
import org.quartz.jobs.ee.jms.SendMsgJob;
import org.quartz.impl.*;
...
JobDataMap dataMap = new JobDataMap();
dataMap.put("jms.connection.factory",
        "ldap://127.0.0.1:1099/Foo");
(*@\colorbox{red!20!green!20!white}{JobDetail job = new JobDetailImpl(); }@*) // UnexpectedException, round 1&2 (*@\label{line:unexcepted}@*)
job.setJobDataMap(dataMap);         
JobExecCtxImpl ctx = new JobExecCtxImpl(job,
(*@\colorbox{red!20}{-     null,         }@*)  //NullPointerException , round 1 (*@\label{line:nullpointer}@*)
(*@\colorbox{green!20}{+     new BaseTrigger()}@*) (*@\label{line:corrected}@*)
);
SendMsgJob instance = new SendMsgJob();
instance.execute(ctx);
\end{lstlisting}
    \caption{The two rounds of exploit generation of CVE-2023-39017 by GPT-4o. The first round includes every line except line ~\ref{line:corrected}. The second round corrects line ~\ref{line:nullpointer} to line ~\ref{line:corrected}.}
    \label{lst:fail_expl}
\end{figure}

The exploit has two fatal issues, one is at line~\ref{line:unexcepted}: the wrong argument of instantiating \code{JobDetailImpl}, and another is at line~\ref{line:nullpointer}: the wrong argument of instantiating \code{JobExecCtxImpl}.
Intuitively, providing feedback to LLM can help it correct its mistakes. 
This turned out to be the right direction, like we provided the error message obtained from exploit execution to GPT-4o, and the regenerated code did fix one error (it changed line~\ref{line:nullpointer} to line~\ref{line:corrected}).
However, it remains challenging for the LLM to identify and rectify its own errors.

In conclusion, generating exploits is challenging, and traditional methods have clear limitations. Although LLMs offer promising potential, leveraging their capabilities to comprehend code and perform AEG remains a non-trivial challenge. In \autoref{sec:approach}, we describe in detail how we address these challenges using a multi-agent framework.

\section{Approach}
\label{sec:approach}

\subsection{Overview}\label{sec:overview}

\begin{figure}[h]
    \centering
    \includegraphics[width=0.47\textwidth]{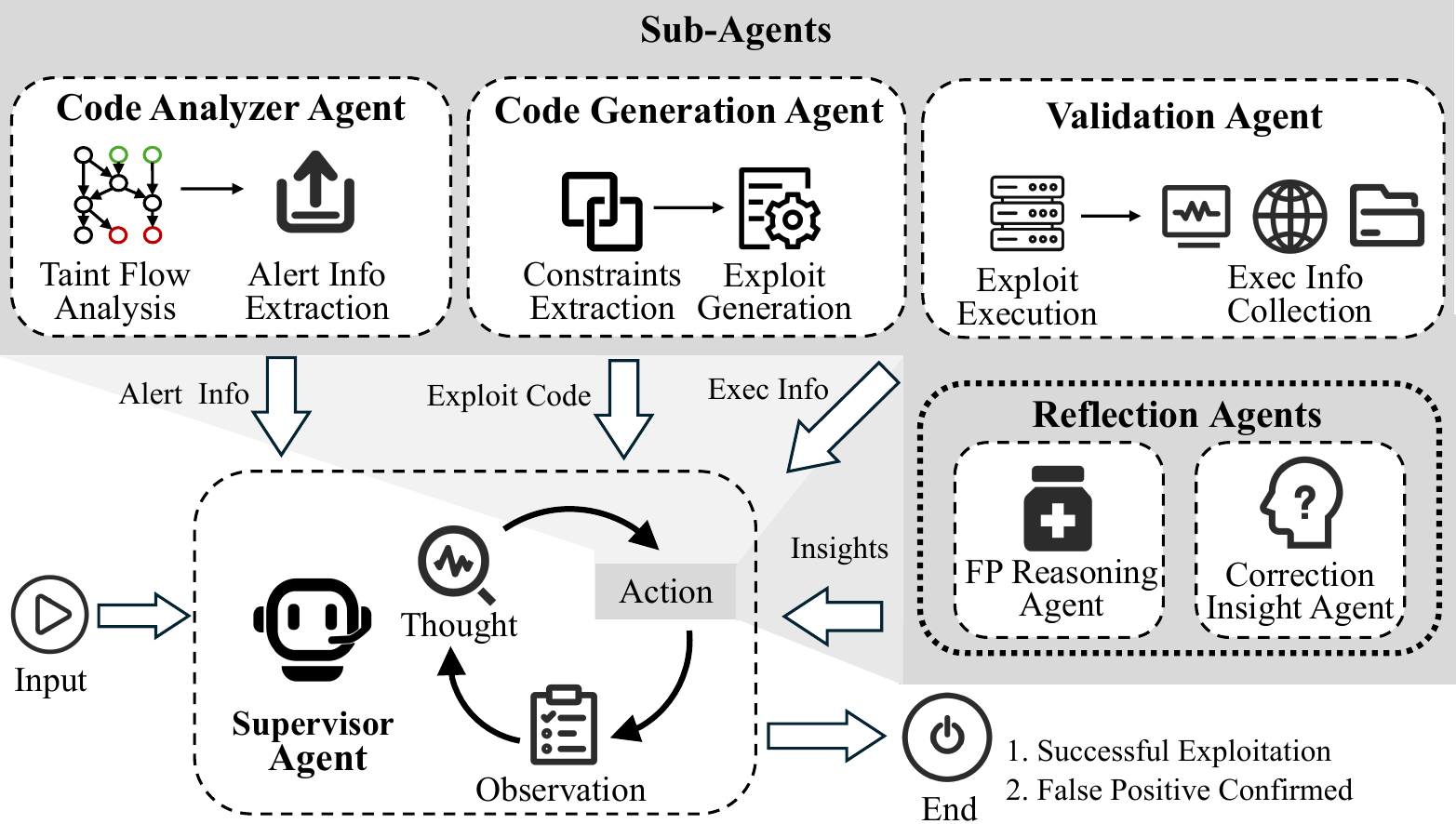}
    \caption{Overview of \vulnsage.}
    \label{fig:vulnsage_main}
            \vspace{-1em}
\end{figure}

Our architecture is based on a multi-agent framework as illustrated in \autoref{fig:vulnsage_main}, motivated by the observation that specialized agents demonstrate superior performance on domain-specific tasks compared to a monolithic LLM~\cite{DBLP:conf/nips/Bo0DFW00W24}, 

One supervisor agent at the top accepts the library code as the input, invokes sub-agents to discover exploitable vulnerabilities with valid exploit code, and for false positive vulnerabilities, it provides detailed reasons.

Sub-agents are responsible for handling specific tasks. For now, we have the following main sub-agents (we omit some subtle agents, such as the install packages agent):
\begin{itemize}[leftmargin=0.5cm, itemindent=0cm,topsep=0.1cm]
    \item The Code Analyzer Agent, for producing the alert information;
    \item The Code Generation Agent, for extracting constraints and producing the exploit code.
    \item The Validation Agent, for executing the exploit code to verify its correctness, and collecting the execution information.
    \item The Reflection Agents, including a False Positive Agent for reasoning whether the alert itself is a false positive, and a Correction Insight Agent for generating \textit{insights} to refine current exploit code (we will illustrate the meaning of \textit{insights} later).
\end{itemize}

In the following illustration, we use Java and JavaScript as examples; however, our approach is generally applicable to a wide range of programming languages.

\subsection{Supervisor Agent}\label{sec:Supervisor}

The Supervisor Agent serves as the core component of \vulnsage. It follows the ReAct (Reasoning and Acting) architecture~\cite{yao2022react, DBLP:journals/corr/abs-2402-14034}.
The ReAct architecture is a prompt design that introduces the available actions within the prompt and guides the LLM to autonomously decide the next action based on previously invoked actions and their observed outcomes, thereby enabling the model to combine reasoning and acting capabilities.

\begin{figure}[h]
    \centering
    \includegraphics[width=0.9\linewidth]{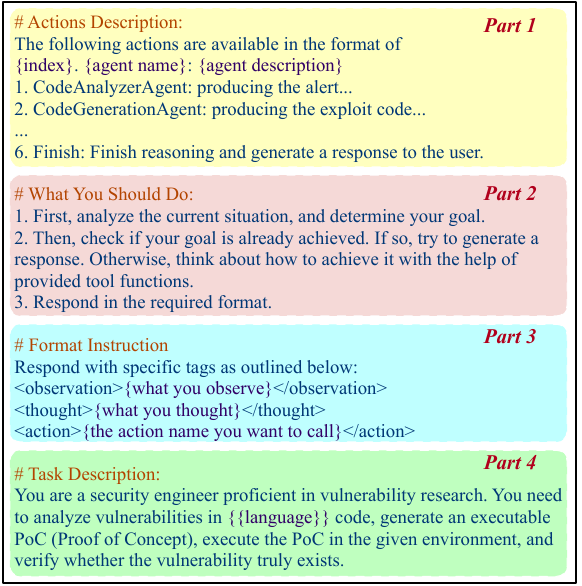}
        \vspace{-0.25cm}

    \caption{The prompt for the Supervisor Agent.}     
    \label{lst:supervisor}
            \vspace{-1em}
\end{figure}

Inspired by AgentScope, an implementation of the ReAct~\cite{DBLP:journals/corr/abs-2402-14034}, we design the prompt as shown in \autoref{lst:supervisor}. The first part lists the optional actions, which are our sub-agents' names and descriptions.
The second and third parts present a three-step reasoning process and format instructions separately to guide LLM to follow the thought-action-observation steps. The last part describes our AEG task.

Note that in our prompt, we do not explicitly enforce the supervisor agent to follow a fixed order of taking actions (e.g., running code analysis first, then code generation, ...). 
The Supervisor Agent autonomously determines itself, as it is guided by "What you should do" in a prompt.
This feature of ReAct is particularly useful for executing the Reflection Agents --- the supervisor will pick one of them after the Validation Agent responds, but not call each of them in order.
Also, this feature lets the Supervisor Agent terminate itself when it calls the finish action once it generates the exploit or concludes that the alert is a false positive.
All of these show the flexibility of the ReAct paradigm, which brings more efficiency to our task.

\subsection{Code Analyzer Agent}\label{sec:StaticAnalysis}
The Code Analyzer Agent aims to provide alerts (the possible vulnerabilities). Each alert contains the vulnerability type, detailed information of dataflow, and an exploit template as \vulnrelated. 
We will explain the meaning of the detailed information of dataflow and the \textit{exploit template} below. \autoref{lst:scan_feeder} shows the agent’s output for the code in \autoref{lst:soure_code_cve}.

\begin{figure}[h]
    \centering
        \begin{lstlisting}[
        basicstyle=\footnotesize\ttfamily, firstnumber=1]
{
 "vulnType": "JNDI.",
 "packageName": "org.quartz-scheduler:quartz-jobs:2.4.0",
 "callChainWithCtx":[ (*@\label{line:callChainWithCtx}@*)
  ["org.quartz.jobs.ee.jms.SendMsgObj.execute ()",
    "SendMsgObj.execute (...) {...}"],
  ["javax.jmx.JmsHelper.context.lookup ()",
    "context.lookup (...) {...}"],
  ...
],
"inputClassSet":[ (*@\label{line:inputClassSet}@*)
"org.quartz.jobs.ee.jms.JobExecCtxImpl._init_ (...){...}",
    ...
],
"sink": "namingCtx.lookup (...) in SendMsgObj.execute",
"entryPoint": "SendMsgObj.execute (JobExecCtx)",
"template": "import org.quartz.jobs.ee.jms; (*@\label{line:template}@*)
SendMsgJob instance=new SendMsgJob(); (*@\label{line:new}@*)
instance.execute (<source>);" (*@\label{line:execute}@*)
}
        \end{lstlisting}
            \vspace{-0.25cm}

\caption{An \vulnrelated of \autoref{lst:soure_code_cve} reported by the Code Analyzer Agent.}     
\label{lst:scan_feeder}
        \vspace{-1em}
\end{figure}

\subsubsection{Taint Flow Information}
We use taint analysis to identify potential vulnerabilities. The entry points are the public methods of the target library’s public classes; sources are the entry points’ parameters; sinks depend on the vulnerability type; and sanitizers include functions like \code{Integer.parseInt} in Java. For each alert, the agent records the following \textit{detailed information}:

\begin{itemize}[leftmargin=0.5cm, itemindent=0cm,topsep=0.1cm]
    \item \callchain (line~\ref{line:callChainWithCtx}), the signature and source code of every function on the taint path from source to sink, which can be directly extracted from the taint-flow results.
    \item $inputClassSet$ (line~\ref{line:inputClassSet}), the set of user-defined class definitions relevant to the input. 
    To construct $inputClassSet$, we start from the parameters of the entry point function and recursively collect all involved user-defined classes, adding each class’s constructor and the constructors of its parent classes into $inputClassSet$, and repeating this process until no new classes are discovered.
\end{itemize}

\subsubsection{Exploit Template}\label{sec:template}

We automatically generate an exploit template (line~\ref{line:template}) for each alert. The template consists of two parts: (1) an import statement of the entry point function; and (2) a call expression that invokes the entry point (line~\ref{line:execute}).

The import statement is trivial to build. The call expression is generated based on the entry point: if the entry point is a function or constructor, we create a call expression that calls the function/class identifier; and if the entry point is a method, we first make an object-creation expression (e.g., line~\ref{line:new}). After that, we make a method call where the object is the one we created. We also introduce a parameter marker to indicate the location of the controllable parameter (i.e., the source).

\subsection{Code Generation Agent}\label{sec:Code Generation Agent}

\begin{figure}[H]
    \centering
    \includegraphics[width=\linewidth]{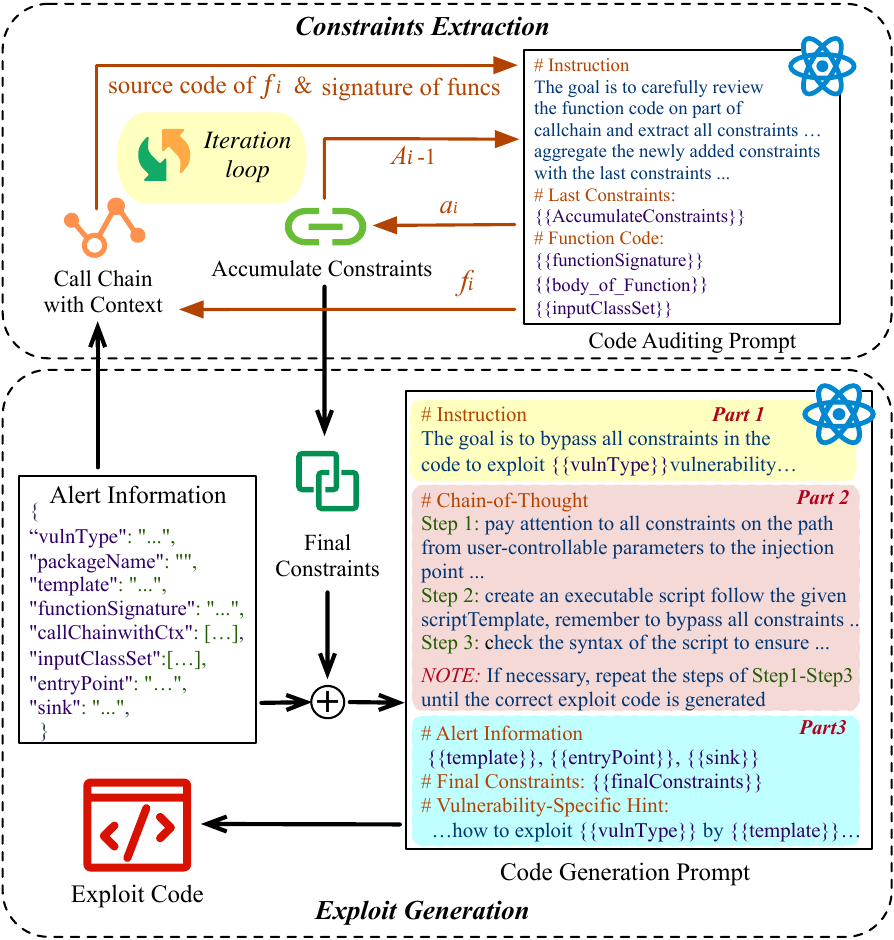}
    \caption{Overview of the Code Generation Agent.}
    \label{fig:prompts}
    \vspace{-1em}
\end{figure}

The Code Generation Agent is shown in \autoref{fig:prompts}, which has two parts: Constraints extraction and Exploit Generation.

\subsubsection{Constraints extraction}\label{codeauditing}
Constraints extraction converts the taint-flow propagation of an alert into a set of constraints. The constraints mean that if the input satisfies these constraints, the input of the entry function can execute code to sink function, as the alert information describes.
Unlike symbolic execution, our constraints are described in natural language. 
For example, from line~\ref{line:JobExecTexImpl}), \autoref{lst:soure_code_cve}, we have a constraint like ``the first argument of \code{JobDetail} must be a class type and the class should be Job class or its subclass''. 
As discussed earlier, since the program logic is complex, LLMs are better at expressing such logic in natural language rather than in a formalized method (because they are trained with large natural language, not languages like smt-lib\footnote{\url{https://smt-lib.org/}}, Rocq\footnote{\url{https://rocq-prover.org/}}, or lean\footnote{\url{https://lean-lang.org/}}). This natural-language representation can also be better understood by other LLM-based agents.

The constraints are extracted incrementally as illustrated in the upper part of \autoref{fig:prompts}.
Let $A_i$ denote the set of constraints accumulated in iteration $i$,  and $f_i$ be the function reviewed in iteration $i$.
Initially, $A_0=\mathit{True}$, $f_0$ is the entry point function.
In each iteration, the prompt is provided with all function signatures in \callchain, $\mathit{inputClassSet}$, the current function code $f_i$, and the previously accumulated constraints $A_{i-1}$.
The LLM then parses $f_i$ into new constraints $a_i$, updates the set of constraints as $A_i = A_{i-1} \wedge a_i$, and returns the pair $\langle f_{i+1}, A_i \rangle$, where $f_{i+1}$ is the next function selected by the LLM for review.
Finally, when the LLM returns $f_i = \varnothing$, it indicates that all relevant functions have been analyzed.

The design offers two advantages. 
First, it splits the large alert information into several shorter prompts, preventing each from exceeding the maximum token limit and allowing the LLM to generate high-quality constraints by focusing on the current function.
Second, it provides LLM the flexibility to skip functions it thinks unimportant, even if they appeared in \textit{\callchain}.

\subsubsection{Exploit Generation}

Exploit Generation is a LLM driven by the ``Code Generation Prompt'' (the lower-right part of \autoref{fig:prompts}), which consists of three parts: instruction, chain-of-thought (CoT), and concrete information:

The instruction (part 1) directs the agent to focus on exploit generation.
The CoT (part 2) contains three few-shot reasoning stages designed to enable the LLM’s ability to perform multi-step reasoning for exploit generation.
Step 3 instructs LLM to check the syntax immediately. Though this check can not guarantee execution success, this improves the quality of the generated exploit code~\cite{DBLP:journals/corr/abs-2503-01307}.
The concrete information(part 3) including $vulnType$, $\mathit{template}$, $\mathit{entrypoint}$, $\mathit{sink}$ and $\mathit{finalConstrains}$ generated from \autoref{codeauditing}. 
Additionally, we provide \code{Vulnerability-Specific Hint} to ensure the exploit code carries the specified payload.
These hints aid validation: for example, for an RCE exploit, the hint requires the exploit code to execute a specific command (e.g. /evalcommand), which the Validation Agent uses to determine whether the exploit succeeds. These hints can be inferred after illustrating \autoref{sec:validation agent}.

\subsection{Validation Agent}\label{sec:validation agent}
The Validation Agent checks whether the exploit succeeds by executing the code in a sandbox. If it fails, it provides runtime feedback from the exploit code execution.

\subsubsection{Validate for exploit code}
In general, the validation executes the exploit code and observes whether the execution matches the oracles.
Different vulnerabilities have different oracles, for example:

\begin{enumerate}[leftmargin=0.5cm, itemindent=0cm,topsep=0.1cm]
    \item \textit{Command Injection} \label{enum:cmdi}: The execution runs a custom command we specify. 
    \item \textit{Code Injection}: The execution executes a code that calls the System Call API to run the custom command. And the call stack contains the library function, not directly from the exploit code.
    \item \textit{Path Traversal}: The execution reads a specified file in the root directory.
    \item \textit{Prototype Pollution}: The execution modifies a specific property in \code{Object.prototype}.
    \item \textit{JNDI Injection}: The execution requests an evil JNDI provider. The provider holds a JNDI object that will invoke the system call to run the command.
\end{enumerate}

To prevent the Code Generation Agent from cheating, the validation pipeline applies both static and dynamic anti-cheat checks. For example, in Command Injection, the Validation Agent performs an analysis and rejects it if it contains direct command-execution API calls (like \code{Runtime.exec}, \code{child\_process.exec}). If such a pattern exists, it returns ``exploit is invalid because it calls command execution API directly: {evidence}''.

\subsubsection{Collection for Executing Information}

During execution of the exploit, the agent collects the \textit{execution information}, including two parts: compilation information and runtime information. 

For compilation information, the agent records all error messages from the compiler if
compilation failures (e.g., syntax issues or missing dependencies), otherwise it records ``compilation success''.

For runtime information, the agent records the execution traces. If there is a runtime error in PoC execution, it also records the error stack. When executing the sink function, it also records the call stack.

\subsection{Reflection Agents} \label{sec:reflection}

Reflection Agents are a set of agents that receive vuln-related information, exploit code, and execution information, reflect on an insight for exploit correction, or a reason for concluding false positives.

\subsubsection{Correction Insight Agent}
\begin{figure}[H]
    \centering
    \includegraphics[width=0.9\linewidth]{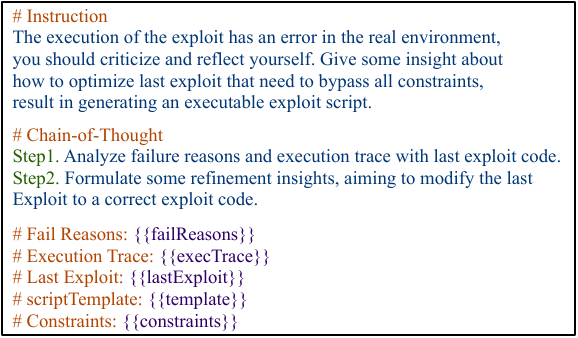}
    \caption{The Prompt for the Correction Insight Agent.}
    \label{fig:reflection_prompt}
        \vspace{-1em}
\end{figure}

This agent is typical for providing an insight for correction, driven by an LLM augmented with a prompt shown in \autoref{fig:reflection_prompt}. 
The prompt with CoT outlines two steps: (1) determining the root cause of execution failures, (2) based on the root cause, reflecting on what other constraints the exploit code should satisfy. These constraints will serve as the agent’s insights and will eventually be incorporated into the $finalConstraints$ illustrated in Code Generation Agent.

\subsubsection{False Positive Reasoning}

This agent reviews the code recorded in the alert and tries to conclude that the current exploit code failure is because the alert itself is a false positive. Since the alert is reported by our static analysis, it concludes the false positive (FP) with the following reasons:

\begin{itemize}[leftmargin=0.5cm, itemindent=0cm,topsep=0.1cm]
    \item Presence of Sanitizer Function: The agent find an unspecified sanitizer function in the middle of the taint flow.
    \item Imprecision of Static Analysis: The agent find that the alert is FP because of the imprecision of call graphs, or some over-approximate modeling of the taint propagation, such as if the taint element flows into an array, the whole array becomes tainted.
\end{itemize}

The reason it returns allows the supervisor to terminate the generation early, which saves the cost and is convenient for the user to confirm false alarms.

\section{Implementation}

Technically, our approach supports all stateless vulnerabilities that can be modeled as taint-flow problems. In our prototype, we support five types of vulnerabilities in Java and JavaScript programs: command injection, code injection, path traversal, prototype pollution, and JNDI injection.

In the Code Analysis Agent, to improve scalability (we analyze both the target code and the dependencies' code), we make several trade-offs on soundness:
To build the call graph, our prototype employs an intra-procedural pointer analysis, which only handles object definition and assignment expressions. 
For each call site, the analysis first attempts to resolve the target using pointer information. 
If this resolution fails, the call is handled differently depending on the language: for Java, we apply Class Hierarchy Analysis, whereas for JavaScript, we select the function with the same name and number of parameters.
Our taint analysis is inter-procedural, context-sensitive, flow-sensitive, and field-sensitive. 
For efficiency, each loop body is analyzed only once per context, and recursive functions are analyzed only if the current context does not contain that function --- a common strategy widely used in static analysis~\cite{10.1145/3192366.3192418}.
For sources and sinks, we collect them from the historical vulnerabilities.

In Validation Agent, for Java, we leverage the Java Native Interface to collect the execution trace.
For JavaScript,  we instrument several sink functions like \code{eval}, \code{child\_process.exec}, etc., to dump the call stack as traces.

For all LLM-driven agents, we use the Qwen3-Max model as it provides the best performance, which will be shown in \autoref{subsec:RQ4}. 
In \autoref{subsec:RQ4}, we also evaluate the performance of GPT-4o, DeepSeek-V31, and Qwen3-Plus. All models use $temperature = 0.0$ and $seed = 1234$ across all experiments.  Among the models listed, Qwen3-Max offers the largest context window at 252K tokens, followed by Qwen3-Plus and GPT-4o (both at 128K), while DeepSeek-V31 supports 96K tokens.
For each alert, \vulnsage is allowed at most 20 attempts to generate a valid result.

We open-sourced our implementation anonymously in \url{https://github.com/Vulnsage/VulnSage}.

\section{Evaluation}\label{sec:evaluation}
To evaluate the effectiveness and efficiency of \vulnsage, we design five research questions.

\begin{enumerate}[label=\textbf{RQ\arabic*}, leftmargin=3em]
    \item\label{enum:rq1} How effective is \vulnsage compared to other tools for exploit generation?
    \item\label{enum:rq2} Can \vulnsage find security vulnerabilities in real-world packages for different languages?
    \item\label{enum:rq3} How much does each component of \vulnsage contribute to the overall effectiveness?
    \item\label{enum:rq4} How much does the choice of LLM affect performance?
    \item\label{enum:rq5} What is the time-consumption and token-consumption cost of \vulnsage?
\end{enumerate}

All experiments are conducted on a server with an Intel (R) Xeon (R) Platinum 8163 CPU @ 2.50GHz (16 cores), 32 GB memory, and Ubuntu 22.04.6 LTS.

For \ref{enum:rq1}, we use a dataset from  \secbench~\cite{hasan2023secbenchjs}, which contains 475 vulnerable server-side npm packages. The vulnerability type we experimented with included code execution, command injection, path traversal, and prototype pollution.

For the rest of our experiment, except \secbench, we collected all packages with their latest stable version from the npm and Maven repositories as dataset \validset~\footnote{The dataset was collected on February 28, 2025}, which contains {59,628} JavaScript and {80,785} Java packages. 

For \ref{enum:rq1}–\ref{enum:rq4}, we measure different techniques and configurations by the number of alerts, proof-of-executable(PoE), vulnerabilities, and exploits, as related work~\cite{marques2025explode} did.
We clarify our terminology as follows:
\begin{itemize}[leftmargin=0.5cm, itemindent=0cm,topsep=0.1cm]
    \item Alert: A finding produced by the program-analysis stage, typically represented as one or more data-flow paths from source(s) to one sink (or an equivalent static warning). 
    \item Vulnerability: The alert is a vulnerability (i.e., true positive) if the paths of alerts are an unintended execution path and the alert is exploitable.
    \item PoE. An automatically generated piece of executable code produced by a technique to prove one of the data-flow paths reported in an alert is executable.
    \item Exploit. A PoE is an exploit if its corresponding alert is confirmed as a vulnerability.
\end{itemize}
For \ref{enum:rq5}, we simply use the running time, the token consumption, and the cost as our metrics.
\subsection{The Result of RQ1}\label{subsec:RQ1}

To show the effectiveness of \vulnsage, we compare \vulnsage with two state-of-the-art approaches ---  \nodemedicfine(NM)~\cite{Darion2025NodemedictFineNdss} and \explodejs(EXPLODE)~\cite{marques2025explode} in \secbench benchmark. The results are in \autoref{tab:secbench-cmp-nm-and-exp}, where ``-'' means the technique does not support the vulnerability type.

\setlength{\tabcolsep}{1.5pt}
\begin{table}[htb!]
    \centering
    \caption{The number of vulnerabilities and Exploits generated by NM, EXPLODE, and \vulnsage. }
    \small
    % \begin{threeparttable}
        \begin{tabular}{@{}lccccccc@{}}
            \toprule
            \multirow{2}{*}{\textbf{VulnType}} 
            & \multicolumn{4}{c}{\textbf{Vulnerabilities}} 
            & \multicolumn{3}{c}{\textbf{Exploits}} \\ 
            \cmidrule(lr){2-5} \cmidrule(lr){6-8}
            & \textbf{Total} 
            & \textbf{NM} 
            & \textbf{EXPLODE} 
            & \textbf{\vulnsage}
            & \textbf{NM} 
            & \textbf{EXPLODE} 
            & \textbf{\vulnsage} \\ 
            \midrule
            \textbf{\pathtraversal}    & 158 & - & 95 & \textbf{151} & - & 82 & \textbf{114} \\ 
            \textbf{\commandinjection}    & 99 & 32 & 56 & \textbf{81} & 26 & 42 & \textbf{78} \\
            \textbf{\codeinjection}    & 32  & 3  & 6  & \textbf{16} & 0  & 3  & \textbf{14}  \\
            \textbf{\prototypepollution}  & 186 & -- & 46 & \textbf{55} & -- & 39 & \textbf{48} \\ 
            \midrule
            \textbf{SUM}     & 475 & 35 & 202 & \textbf{303} & 26 & 166 & \textbf{254} \\ 
            \bottomrule
        \end{tabular}
    % \end{threeparttable}
    \label{tab:secbench-cmp-nm-and-exp}
\end{table}
\setlength{\tabcolsep}{6pt} 

We can see that for all of the vulnerability types, \vulnsage performs better than the others. 
And in the end, \vulnsage successfully generated 88 (34.64\%) more exploits compared to EXPLODE and 228(89.76\%) more exploits than NM.
    
The ``Total'' column in vulnerabilities indicates the total number of vulnerabilities in \secbench.
From the vulnerability metrics, we observe that \vulnsage's Code Analyzer Agent detects more vulnerabilities than the other two tools' detections, giving \vulnsage an initial advantage. 
NM cannot provide as many vulnerabilities because NM uses dynamic taint analysis to detect vulnerabilities; therefore, its execution does not cover all parts of the code.
EXPLODE uses static analysis to find vulnerabilities, but it reports the vulnerability only when the analysis can successfully generate constraints for the execution path of vulnerabilities.

But this raises a follow-up question: does \vulnsage generate more exploits merely due to the advantages of its static analysis?
To answer this question, we construct \autoref{tab:secbench-cmp-nm-and-exp-fair}, where we extract the vulnerabilities that are commonly detected by all three tools in the first-stage analysis and evaluate how many exploits each tool can successfully generate (because NM does not support path traversal or prototype pollution, we extract, for these two types, only the vulnerabilities that are jointly detected by \explodejs and \vulnsage).
From the last column, we can see that \vulnsage still performs better than others, because across all kinds of vulnerabilities, \vulnsage misses only 3, whereas NM misses 5 and EXPLODE misses 19.
Moreover, there are no cases that NM or EXPLODE can generate, but \vulnsage cannot. After manually examining the failed cases, reasons why \vulnsage successfully generates exploits, but NM or EXPLODE fail are as follows:
\begin{enumerate}[leftmargin=0.5cm, itemindent=0cm,topsep=0.1cm]
    \item \vulnsage excels at handling complex string constraints.
    In many exploits, specific strings are crucial in determining whether the execution flow reaches the sink and whether the sink function actually triggers the attack payload.
    Programs often compare these strings using regular-expression pattern matching or parse the string according to a particular syntax (e.g., the string should be in JSON, XML, or follow the grammar of another programming language). We will present an example in the case study for RQ2.
    \item \vulnsage also outperforms in constructing complex objects and functions as valid inputs, as we illustrate in \autoref{sec:MOTIVATING}.
    Some exploits require obtaining the specific function via the prototype chain (e.g., \code{''.constructor.constructor}) to bypass the check, or passing a particular function (e.g., \code{require} or \code{module}) or a previously defined class instance as inputs. 
\end{enumerate}

All of those constraints are either hard to solve or even impossible to describe in the current SMT language. While for fuzzing, it is hard to cover all possible inputs.

We check 3 cases that we miss. 
They either require system-level packages installed in advance (\vulnsage is not allowed to install system-level packages for a fair comparison) or the exploit must run from the command line.

\setlength{\tabcolsep}{1.5pt}
\begin{table}[tb]
    \caption{Exploits generated by NM, EXPLODE, and \vulnsage for vulnerabilities that were detected by the static analyses of all three tools.}
        \vspace{-0.25cm}

    \label{tab:secbench-cmp-nm-and-exp-fair}
    \small
    \begin{tabular}{@{}ccccc@{}}
        \toprule
        \multirow{2}{*}{\textbf{VulnType}} & \textbf{Vulnerabilities} & \multicolumn{3}{c}{\textbf{Exploits}}             \\ \cmidrule(l){3-5} 
        &      \textbf{$\text{NM} \cap \text{EXPLODE} \cap \text{\vulnsage}$}              & \textbf{NM} & \textbf{EXPLODE} & \textbf{\vulnsage} \\ \midrule
        \textbf{\pathtraversal}                & 94                     & -             & 82         & \textbf{91}        \\
        \textbf{\commandinjection}                & 24                     & 20             & 21         & \textbf{24}        \\
        \textbf{\codeinjection}               & 1                      & 0              & 0          & \textbf{1}         \\
        \textbf{\prototypepollution}             & 18                     & -              & 15         & \textbf{18}        \\ \midrule
        \textbf{Total }                & 137                     & 20             & 118         & \textbf{134}        \\ \bottomrule
    \end{tabular}
    \vspace{-2em}
\end{table}
\setlength{\tabcolsep}{6pt}

Moreover, both \autoref{tab:secbench-cmp-nm-and-exp} and \autoref{tab:secbench-cmp-nm-and-exp-fair} show that \vulnsage performs better not only on specific vulnerability types but across all of them. Therefore, we expect it will continue to perform well when extended to other vulnerability categories.

In summary, \vulnsage can generate more successful exploits than other tools because of its more sound static analysis and well-designed LLM-based exploit generation framework.
\subsection{The Result of RQ2}\label{subsec:RQ2}

To evaluate \vulnsage's performance in the real world, we run it on \validset. The results are shown in \autoref{tab:0day}.

\setlength{\tabcolsep}{6pt}
\begin{table}[htb]
    \centering
    \small
    \caption{\vulnsage performance on \validset.}
        \vspace{-0.25cm}
    \label{tab:0day}
    % \begin{threeparttable}
        \begin{tabular}{@{}cccccc@{}}
            \toprule
            \textbf{Language}         & \textbf{VulnType} & \textbf{Alerts} & \textbf{PoEs} & \textbf{Exploits} & \textbf{0-day} \\ \midrule
            \multirow{3}{*}{\textbf{JavaScript}}   & \commandinjection          & 1,429              & 235              & 58                 & \textbf{43}             \\
            & \codeinjection          & 173                & 29               & 25                 & \textbf{15}              \\
            & \prototypepollution       & 160                & 59               & 59                 & \textbf{48}             \\
            & \pathtraversal          & 288                & 196               & 7                 & \textbf{7}             \\ \midrule
            \multirow{2}{*}{\textbf{Java}} & \commandinjection          & 191                & 66               & 19                  & \textbf{19}              \\
            & \jndiinjection          & 172                & {87}               & {15}                 & \textbf{14}             \\ \midrule
            \multicolumn{2}{c}{\textbf{Total}}               & 2,413              & 672              & 183                 & \textbf{146}             \\ \bottomrule
        \end{tabular}
    % \end{threeparttable}
\end{table}
\setlength{\tabcolsep}{6pt}

Based on 2,413 alerts, \vulnsage successfully generates PoE for approximately 27.85\%(672/2,413) of them. Among 678 PoEs, 183 target actual vulnerabilities.
We randomly check 60 alerts, and conclude that 38.33\%(23/60) of the alerts belong to vulnerabilities.
And for vulnerabilities, \vulnsage can generate exploits for 56.52\%(13/23) of them.
After excluding the vulnerabilities that have been discovered before, 146 of 183 vulnerabilities are new (\textit{0-day} vulnerabilities).
We have reported these 0-day vulnerabilities to the CVE team. Currently, 73 have been assigned CVE IDs, and 22 have been disclosed.\footnote{We will disclose the vulnerabilities with assigned CVE IDs upon publication of this paper.}

We ignore alerts that are consistent with the library’s intended behavior and randomly investigate 40 alerts for which \vulnsage failed to generate exploits.
Among them, (1) 52.5\% are actually false positives (FPs), mainly because the inputs are sanitized before reaching the sink or due to imprecise taint analysis.
(2) Another 22.5\% fail because of missing system-level packages, as discussed in RQ1.
(3) The rest are because the input needs to satisfy complex constraints, such as some libraries need an input that is an ASCII encoded string representing a PNG image.
In these cases, we find that \vulnsage repeatedly mutates an incorrect exploit candidate in every iteration, even though that candidate is doomed to fail.
If we let \vulnsage restart from scratch, it may nevertheless eventually produce a working exploit. 

For the first two cases, \vulnsage correctly reports all the failure reasons,
which means, in addition to confirming true positives by generating exploits, \vulnsage also reduces the time required to validate false positives.

In conclusion, VulnSage generates 183 exploits from 2,413 alerts, including 146 zero-day vulnerabilities. Among these, 73 have been assigned CVE identifiers, and 23 have been publicly disclosed\footnote{As of January 28, 2026}. For the failed cases, \vulnsage precisely explains the underlying reasons, enabling users to efficiently eliminate false positives from static analysis.

To validate these 0-day vulnerabilities, we evaluated EXPLODE on 113 cases. Our evaluation shows that EXPLODE only successfully generated 1 PoC. Failures fall into two categories: (1) inability to build a constraint-solving template because of unmodeled JavaScript features, and (2) failure to solve complex constraints.

\paragraph{Case Study} \autoref{lst:case-study-a} shows a code-injection vulnerability for which \vulnsage successfully generated an exploit(at line~\ref{case:exploit}). We omit the implementation of \code{esprima.parse(str)} (which parses \code{str} into a JavaScript AST), \code{traverse(ast, func)} (which traverses each node in \code{ast} and applies the callback \code{func}), and \code{nodeToString(ast, node)} (which extracts the source string corresponding to \code{node} from the AST). 
The sink is the \code{eval} call on line~\ref{case:sink}; its input is conditionally controlled by the external \code{fileContents}. Line~\ref{case:parse} constraints that the exploit payload must be a JavaScript string. From the conditions at line~\ref{case:arg}, the parsed AST must contain an \code{ExpressionStatement} whose sub-node satisfies the checks performed at lines~\ref{case:check-begin}–\ref{case:check-end}. These constraints are difficult for SMT solvers and fuzzers to satisfy directly. 
By contrast, \vulnsage successfully generates inputs that meet these constraints in only a few iterations.

\begin{figure}[htbp]
    \centering
    \begin{lstlisting}[language=JavaScript, firstnumber=1]
exports.findConfig = function (fileContents) {
  var jsConfig, foundConfig,
    astRoot = esprima.parse(fileContents);  (*@\label{case:parse}@*)
  traverse(astRoot, function(node) {
    var arg;
    if (node.type === 'CallExpression' && c.type === 'Identifier' && c.name === 'require') {
      // ... 
    } else {
      arg = getRequireObjectLiteral(node);
      if (arg) (*@\label{case:arg}@*)
        jsConfig = nodeToString(fileContents, arg);
    }
    if (jsConfig)
      foundConfig = eval('(' + jsConfig + ')'); // Sink here (*@\label{case:sink}@*)
  });
}

function getObjectLiteral(node) {
  if (node.type && node.type === 'ExpressionStatement' && (*@\label{case:check-begin}@*)
      (node.expression && !node.arguments) &&
      (node.expression.type && node.expression.type === 'ObjectExpression') &&
      (node.expression.properties && node.expression.properties.length > 0) &&
      (node.expression.properties[0].type && node.expression.properties[0].type === 'Property')) (*@\label{case:check-end}@*)
    return node;
};

// exploits:
findConfig("({\"paths\":{\"a\":(require('child_process').exec('touch flag'),1)}})") (*@\label{case:exploit}@*)
    \end{lstlisting}
        \vspace{-0.25cm}
    \caption{A 0-day vulnerability discovered by \vulnsage.}
    \label{lst:case-study-a}
\end{figure}

\subsection{The Result of RQ3}\label{subsec:RQ3}
We evaluate the following \vulnsage configurations:
\noindent
\begin{enumerate}[leftmargin=0.5cm, itemindent=0cm,topsep=0.1cm]
    \item \textbf{Full} : \vulnsage with all components.
    
    \item \textbf{MiniAlert}: 
    We do not supply taint-flow details (i.e., ``\callchain'') in vulnerability information; instead, we provide the LLM with the repository’s complete source code.
    
    \item \textbf{NoRefl}: \vulnsage without Reflection Agents introduced in \autoref{sec:reflection}, meaning that the exploit code is generated by a one-shot LLM.

    \item \textbf{NoTrace}: \vulnsage disables execution path tracking, as described in \autoref{sec:validation agent} as feedback. In this mode, the feedback only contains runtime errors.
\end{enumerate}

The results of these configurations are shown in \autoref{tab:ablationstudy}, where coverage(Cov.) is calculated as the number of exploits generated by each configuration divided by the number generated in full mode.

\setlength{\tabcolsep}{4pt}
\begin{table}[tbph]
\caption{Exploits generated across different configurations.}\label{tab:ablationstudy}
\vspace{-0.25cm}
\small
\begin{tabular}{@{}ccccccc@{}}
\toprule
\textbf{}                     & \multicolumn{4}{c}{\textbf{JavaScript}}                                                & \multicolumn{2}{c}{\textbf{Java}}       \\ \cmidrule(r){2-5} \cmidrule(r){6-7} 
\multicolumn{1}{c}{\textbf{}} & \multicolumn{2}{c}{\secbench} & \multicolumn{2}{c}{{\validset}}        & \multicolumn{2}{c}{\validset} \\ \cmidrule(r){2-3} \cmidrule(r){4-5}\cmidrule(r){6-7} 
                              & {Exploits}    & {Cov.}   & \multicolumn{1}{c}{{Exploits}} & {Cov.} & {Exploits}              & {Cov.} \\ \midrule
\textbf{Full}                 & 242           & 100\%         &  519       & 100\%      & 153                      & 100\%      \\
\textbf{MiniAlert}             & 216	            & 89.26\%         & 295	                            & 56.84\%      & 73                      & 47.71\%       \\
\textbf{NoRefl}             & 167            & 69.01\%         & 352                            & 67.82\%       & 51                      & 58.82\%       \\
\textbf{NoTrace}             & 206             & 85.12\%               & {417}          & {80.35\%}             & {121}  & 79.08\%  \\ \bottomrule
\end{tabular}
\end{table}
\setlength{\tabcolsep}{6pt}

The results indicate that all parts affect the usability of \vulnsage,  with \textbf{NoTrace} having a relatively smaller impact than the first two.

\textbf{MiniAlert} is the most influential factor in terms of impact.
First, its strong performance on \secbench can be attributed to the fact that the dataflows of many vulnerabilities are in a single file.
For real-world cases, we examine {57\%} instances missed by \textbf{MiniAlert}. Among them, {10.26\%} are missed because the prompt exceeds the LLM's maximum context size, while the remaining cases are missed because the LLM becomes overwhelmed by the large codebase and fails to identify the vulnerability-related code for reasoning.
This is understandable, as LLMs generally struggle with complex dataflow. 
Rather than expecting the LLM to infer such relationships on its own, it is more effective to provide it with results from static analysis.

We further analyze 30\% of failed cases in \textbf{NoRefl}. 
The primary reason is that the LLM often fails to generate correct inputs to invoke the entry point in a single attempt, or the generated exploit contains compilation errors.
This observation is consistent with the phenomenon discussed in \autoref{sec:MOTIVATING}.
In \textbf{Full} mode, the LLM can typically resolve compilation errors within 1-3 rounds. Correcting issues related to incorrect arguments within approximately 5 rounds.
On average, exploits that failed in the first round require 8 rounds of back-and-forth to correct.

\textbf{NoTrace.} is the last important thing. Because even without that part of the information, Reflection Agents still work with compile-time and runtime error messages.
It is a good sign because collecting this part requires significant engineering work. Considering the user wants to extend our tools to other languages. Even without this part, \vulnsage can still keep 81.40\%($(206+417+121)/(242+519+153)$) of its usability.

In summary, each component of \vulnsage contributes to its overall effectiveness, with the detailed vulnerability information being the most critical, followed by reflection agents and execution path tracking.
\subsection{The Result of RQ4}
In this section, we replace our default LLM (Qwen3-Max) with others to study how the LLM affects our approach. \autoref{tab:model-change} shows the result on \secbench:

\setlength{\tabcolsep}{3.5pt}
\begin{table}[tbh]
    \centering
    \caption{Exploits generated using different models.} %\wenyuan{cwe to vulntype}}
    \label{tab:model-change}
    \vspace{-0.25cm}
    \small
    % \begin{threeparttable}
        \begin{tabular}{@{}lccccc@{}}
            \toprule
\textbf{Model} & \textbf{\pathtraversal} & \textbf{\commandinjection} & \textbf{\codeinjection} & \textbf{\prototypepollution} & \textbf{Total} \\
\midrule
\textbf{GPT-4o@{24-11-20}} & 110 & 64 & 8 & 30 & 212\\
\textbf{DeepseekV31@{25-08-21}} &  104 & 77 & 13 & 45 & 239\\
\textbf{Qwen3-Plus@{25-04-28}} & 106 & 68 & 7 & 41 & 222 \\
\textbf{Qwen3-Plus@{25-09-11}} & 108 & 74 & 11 & 44 & 237\\
\textbf{Qwen3-Max@{25-09-24}} &  \textbf{114} & \textbf{78} & \textbf{14} & \textbf{48} &\textbf{254}\\
            \bottomrule
        \end{tabular}
    % \end{threeparttable}
\end{table}
\setlength{\tabcolsep}{6pt}

The results show that Qwen3-Max is the best model, we estimate it is 
because Qwen3-Max introduces more advanced architecture, which is beneficial for solving complex, multi-step tasks\cite{qiu-etal-2025-demons}.
Moreover, models released in the same period(e.g., DeepseekV31, Qwen3-Plus@25-09-11, and Qwen3-Max) show similar performance in exploit generation, which is consistent with the previous research~\cite{DBLP:conf/acl/PengYDZZZGZ25}.
Whereas newer models tend to outperform older ones, the comparison between Qwen3-Plus@25-04-28 and Qwen3-Plus@25-09-11 clearly confirms this observation, as the two are almost identical in most aspects except that Qwen3-Plus@25-09-11 is trained on a newer dataset.
This suggests that users will benefit from improved performance for free as more powerful models are released in the future.

In conclusion, different models do affect the results; newer models perform better, and Qwen3-Max currently is the best.
\subsection{The Result of RQ5}\label{subsec:RQ4}
\autoref{tab:efficiency} summarizes the average time and token costs of \vulnsage.
The \emph{input} means the number of tokens in the prompt (input to the LLM), and the \emph{output} means the number of tokens generated by the LLM as output.
For cases,
\textit{Succ.} denotes the cases for which \vulnsage generates a successful exploit, while \textit{Failed.} denotes \vulnsage failed to generate an exploit.
\textit{All.} represents the aggregation of both successful and failed cases.
The last line (\textit{All / All}) indicates the overall average across all cases in all languages.

\setlength{\tabcolsep}{6pt}
\begin{table}[tbh]
    \centering
    \caption{Execution cost of \vulnsage across different tasks.}
    \label{tab:efficiency}
    \vspace{-0.25cm}
    \small
    % \begin{threeparttable}
        \begin{tabular}{@{}lccccc@{}}
            \toprule
            \textbf{Language / Case} & \textbf{Time(s)} & \textbf{Input} & \textbf{Output} & \textbf{cost(\$)}\\ 
            \midrule
\textbf{JavaScript / Succ.} & 231.08          & 553,877        & 21,674 & 0.53\\
\textbf{JavaScript / Failed}      & 646.32 & 969,542 & 40,941 & 0.93\\
\textbf{JavaScript / All}   & 348.32 & 670,242 & 27,114& 0.64\\
            \midrule
\textbf{Java / Succ.} & 262.33 & 988,016 & 43,896 & 0.96\\
\textbf{Java / Failed}     & 567.99 & 1,092,088 & 52,907 & 1.07  \\
\textbf{Java / All}   & 480.45 & 1,062,282 & 50,326 & 1.04\\
            \midrule
\textbf{All / All} & 458.17 & 996,165 & 46,412 & 0.97 \\
            \bottomrule
        \end{tabular}
    % \end{threeparttable}
\end{table}
\setlength{\tabcolsep}{6pt}

The last line indicates that each alert takes approximately 458.17 seconds (about 8 minutes) and costs 996,165 input tokens and 46,412 output tokens, which is acceptable for practical use. Under Qwen's October 2025 pricing(\$0.00082 per thousand Input token and \$0.00329 per thousand output token), each vulnerability costs \$0.97.
In the worst-case scenario, the cost is \$3.49. 
We manually inspect this case and confirm that the reported vulnerability is a false positive due to the presence of a sanitizer. 
Our approach takes 10 iterations to reason about it, and we estimate that an expert will need more than one hours to perform the same reasoning without our method. 
Moreover, the cost of Qwen3-Plus@25-09-11 is \$0.2.
Considering it performs close to Qwen3-Max, it may also be a good choice in practice.

We can see that the cost is higher in Java than in JavaScript.
Because Java code is often more verbose, the dataflow for vulnerabilities in Java is more complex, and implementing the same functionality (including an exploit) usually requires more lines of code. 
For example, we find that for Java, the related function in \callchain consumes 2,586 tokens on average, while JavaScript takes 1,191 tokens.

In conclusion, \vulnsage has effective performance in generating exploit code, with 8 minutes and \$0.97 per vulnerability.

\section{Threats to Validity} 

The dataset inevitably affects the exploit generation results.
In \autoref{subsec:RQ1}, we make our best effort to identify suitable datasets. However, we are unable to find a suitable one for Java.
Nevertheless, by comparing our approach with two traditional techniques, we can already observe the \vulnsage's strengths (and weaknesses). These observations are general.
Moreover, our evaluation uses a large number of real-world applications, which minimizes this potential threat as much as possible.

The LLM we use may have memorized the exploit in SecBench.js, which may threaten the result we get in RQ1. In RQ2, we used the real-world dataset, and the discovered zero-day vulnerabilities demonstrate that our method’s effectiveness is not coming from memorization.

The exploit generation process depends on the alerts provided by the code analyzer. 
Different analyzers may produce alerts of varying quantity and quality, which can, in turn, affect the experimental results.
Table~\ref{tab:secbench-cmp-nm-and-exp-fair} minimizes the differences as much as possible and shows \vulnsage still performs better.
Furthermore, we believe that our architecture can integrate results from other static analyzers, which we consider a promising direction for future work.

We currently evaluate only 5 types of vulnerabilities and 2 programming languages. 
The performance may vary for other kinds of vulnerabilities or programming languages. 
However, as demonstrated in \autoref{subsec:RQ1}, the advantage of \vulnsage lies in its fundamental design rather than in the particular vulnerabilities or languages under study. 
We therefore expect that evaluating vulnerabilities detected by similar static analysis techniques or across different programming languages would not change our overall conclusions.
\section{Related Work}

Considering that, in a broad sense, AEG aims to precisely discover program vulnerabilities, we will discuss our related work from the perspective of automated vulnerability detection, which can be categorized into static and dynamic analysis. Static analysis can be further divided into abstract interpretation-based approaches and symbolic execution. 

The abstract interpretation-based~\cite{DBLP:conf/popl/CousotC77} approaches, such as taint analysis~\cite{DBLP:journals/cacm/Denning76,DBLP:conf/sp/KangXLGHVC23,DBLP:conf/icse/ZhongLWDSLL23}, often pursue a sound analysis --- they can find potential vulnerabilities as much as possible. 
However, this often results in a large number of false positives, which in turn motivates the use of automatic exploit generation (AEG) techniques for validation. 
Our code analysis also falls into this category, and in \ref{enum:rq1}, we have shown the advantage of this choice.

Symbolic execution~\cite{DBLP:journals/cacm/King76}(SE) detects vulnerabilities by solving constraints abstracted from program execution, which is inherently capable of producing concrete inputs that demonstrate the existence of vulnerabilities.
We categorize \explodejs ~\cite{marques2025explode} into this class, although it employs a sophisticated dataflow analysis to generate constraints as a preliminary step.
As demonstrated in \ref{enum:rq1}, SE-based techniques often miss true positives because real-world code often involves higher-order constraint reasoning that current SMT solvers cannot handle automatically.
To mitigate this limitation, several studies such as KLEE~\cite{cadar2008klee, DBLP:conf/icics/CorinM12} and {ANGR}~\cite{DBLP:conf/sp/Shoshitaishvili16} combine SE with dynamic analysis, execute the program concretely to reason about higher-order behaviors rather than relying solely on SMT solving.
Nevertheless, despite their practicality, these hybrid approaches generally remain less sound than analyses based on abstract interpretation.

Dynamic analysis, such as fuzzing~\cite{DBLP:journals/cacm/MillerFS90} and dynamic taint analysis~\cite{DBLP:conf/ndss/NewsomeS05}, tests the program by actually executing it. During program execution, the analysis applies oracles at runtime to detect misbehavior.
\nodemedicfine~\cite{Darion2025NodemedictFineNdss} is built on top of this, which first finds vulnerability by dynamic taint analysis, then refines the input using SMT to synthesize the exploit.
However, dynamic analysis continues to struggle with two long-standing limitations. First, achieving adequate code coverage is difficult, which reduces the likelihood of reaching deep or rare program behaviors. Second, executing interpreted languages (e.g., JavaScript) or languages with a VM runtime (e.g., Java) imposes substantial performance costs~\cite{DBLP:conf/ccs/KerstenLP17}. Both issues undermine the efficiency and effectiveness of dynamic approaches, as shown by our \ref{enum:rq1} results.

As LLMs become widely adopted, researchers are actively exploring their applications to vulnerability discovery and automated exploit generation. Magneto~\cite{zhang2024magneto}, QLPro~\cite{li2024qlpro},\etal{Wang}~\cite{wang2025vulagenthypothesisvalidationbasedmultiagent}, and \etal{Li}~\cite{DBLP:journals/corr/abs-2505-13452} respectively leverage LLMs to enhance the capabilities of fuzzing, static analysis, and SMT solving.
However, these works do not fundamentally overcome the intrinsic limitations of those underlying techniques, as discussed above.
\etal{Jin}~\cite{2505.01065} and \etal{Fang}~\cite{fang2024llm} study the possibility of using LLM and LLM agents to generate exploits and demonstrate that without any additional information, LLMs perform poorly in exploit generation. This observation motivates our design.
PwnGPT~\cite{DBLP:conf/acl/PengYDZZZGZ25} also uses LLM to improve AEG. It only works on simple and small programs used in CTF and fails to exploit real-world programs that have complex code logic.
PocGen~\cite{erbayrak2024pocgen} generates exploit code using disclosed CVE information, but it cannot generate exploits for undisclosed ones.
Wang and Zhou~\cite{DBLP:journals/corr/abs-2508-21579} propose a multi-agent framework for generating PoCs for Android vulnerabilities; it is unclear whether their technique generalizes to package-level vulnerabilities. Additionally, their framework also cannot handle long contexts, which degrades exploit correctness when dataflow spans multiple files or long call chains. In contrast, our approach addresses this limitation by accumulating and summarizing the constraints underlying long dataflow.
\etal{Nitin}~\cite{DBLP:journals/corr/abs-2507-15241} propose a multi-agent framework for generating exploits targeting Java, C, and C++ projects. However, their approach does not incorporate static analysis to identify vulnerabilities; instead, the LLM is provided with the entire project as input. Moreover, when exploit generation fails, the LLM receives no deep feedback such as the execution-path information that our framework supplies for reasoning. As demonstrated by the results of \textbf{MiniAlert} and \textbf{NoTrace} in \ref{enum:rq3}, omitting such key inputs --- especially information derived from static analysis --- significantly reduces the success rate of exploit generation.
\etal{Zhu}~\cite{DBLP:journals/corr/abs-2406-01637} employ a multi-agent framework to exploit vulnerabilities of real-world web applications . However, their technique is closer to dynamic analysis---the LLM generates exploits purely by interacting with web pages and observing the feedback. This setting differs from our target scenario, and we have discussed the disadvantage of dynamic analysis.

\section{Conclusion}
In this paper, we discuss the limitations of traditional automated exploit generation techniques and the challenges of applying LLMs to this task. We introduce \vulnsage, a multi-agent framework that leverages the power of LLMs for automated exploit generation.
The experimental results demonstrate that our approach can generate 34.64\% more exploits than the state-of-the-art and discover 146 0-day vulnerabilities in real-world packages. 
We encourage future work to investigate more efficient agent architectures and to further explore the potential of LLMs in AEG.

\newpage
\bibliographystyle{ACM-Reference-Format}
\bibliography{ref}

\end{document}